\documentclass[useAMS,usenatbib]{mnras}
\usepackage{graphicx}
\usepackage[none]{hyphenat}
\usepackage{comment}
\usepackage[normalem]{ulem}

\def\msun{{\,{\rm M}_\odot}}
\def\mearth{{\,{\rm M}_\oplus}}

\def\simlt{\lower.5ex\hbox{$\; \buildrel < \over \sim \;$}}
\def\simgt{\lower.5ex\hbox{$\; \buildrel > \over \sim \;$}}

\title[Clump formation in the Galactic Centre] {Clump formation through colliding stellar winds in the Galactic Centre}
\author[D. Calder\'on et al.]{D. Calder\'on$^{1}$\thanks{E-mail: dcaldero@astro.puc.cl}, A. Ballone$^{2,3}$, J.
Cuadra$^{1}$, M. Schartmann$^4$, A. Burkert$^{2,3}$ and \newauthor S. Gillessen$^{2}$\\
$^1$Instituto de Astrof\'isica, Facultad de F\'isica, Pontificia Universidad Cat\'olica de Chile, 782-0436 Santiago, Chile\\
$^2$Max Planck Institute for Extraterrestrial Physics, P.O. Box 1312, Giessenbachstr., D-85741 Garching, Germany\\
$^3$Universit\"atssternwarte der Ludwig-Maximilians-Universit\"at, Scheinerstr. 1, D-81679 M\"unchen, Germany\\
$^4$Centre for Astrophysics and Supercomputing, Swinburne University of Technology, Hawthorn, Victoria, 3122, Australia}
\begin{document}

\label{firstpage}

\date{Draft \today}

\pagerange{\pageref{firstpage}--\pageref{lastpage}} \pubyear{2015}
\maketitle

\begin{abstract}
	The gas cloud G2 is currently being tidally disrupted by the Galactic Centre super-massive black hole, Sgr A*. The region 
	around the black hole is populated by $\sim 30$ Wolf-Rayet stars, which produce strong outflows.  We explore 
	the possibility that gas clumps, such as G2, originate from the collision of  stellar winds via the \textit{non-linear thin shell instability}. 
	Following an analytical approach, we study the thermal evolution of slabs formed in the symmetric collision of winds, evaluating whether instabilities occur, and 
	estimating possible clump masses. We find that the collision of relatively slow ($\simlt 750$ km s$^{-1}$) and strong ($\sim 10^{-5}$ $\msun$ yr$^{-1}$) 
	stellar winds from stars at short separations ($<10$~mpc) is a process that indeed could produce clumps of G2's mass and above.  
	Such short separation encounters of single stars along their known orbits  are not common in the Galactic Centre, making this process a possible
	but unlikely origin for G2. We also discuss clump formation in close binaries such as IRS 16SW and in asymmetric encounters as promising 
	alternatives that deserve further numerical study.  
\end{abstract}


\section{Introduction}
	\label{sec:intro}
	\cite{G12} detected a moving diffuse object, the so-called G2 cloud, on its way towards Sgr A*, the radio source identified as the 
	central massive black hole of our Galaxy \citep[see][for a review]{Genzel10}. The tidal disruption of this cloud is being monitored 
	by different groups \citep[e.g.,][]{Eckart13, G13, Phifer13} and provides a unique opportunity to test accretion physics, due to both 
	its proximity and the timescale on which it happens. \cite{G12} estimated the mass of G2 to be $\sim 3 \mearth$ from its line emission. 
	However, the nature of G2 has not been clarified yet. In particular, there is an ongoing debate on whether the diffuse cloud contains a 
	compact mass (likely a star). \cite{W14} presented the detection of a compact source at 3.8 $\mu$m (thermal dust emission) that would 
	correspond to G2 during pericentre passage. Its survival as a compact source to the close passage ($\sim 2000\,$Schwarzchild radii) 
	suggests the existence of a central star keeping it bound, which they argue is a binary merger product \citep{Prodan15}.Other possible 
	explanations for a central mass in G2 include  an evaporating proto-planetary disc or the wind of a T-Tauri star  \citep{M12,S13,B13,D14}.
	
	On the other hand, \cite{P15} argued in favour of a purely gaseous cloud nature for G2 using their Brackett-$\gamma$ observations. 
	They interpreted this source as a bright knot of a larger gas streamer that includes a G2-type object called G1 in a similar orbit, 
	but preceding it by 13 yr. G1 and G2 could be explained as the result of the partial tidal disruption of a star \citep{Guillochon14} 
	or as one of many gas clumps created by the collision of stellar winds from the young stars in the Galactic centre \citep{B12,S12}. 
	Such dense, cold clumps are copiously produced in the Smoothed-Particle Hydrodynamics (hereafter SPH) 
	simulations of the Galactic centre gas dynamics performed by \cite{C05,C06,C08,C15} (see also  \citealt{Luetzgendorf15} \S~3.3), 
	and could survive pericentre passage if magnetised \citep{McCourt15}. 
	Moreover, G2's orbit lies on the plane of  the `clockwise disc', defined by the orbits of many young stars \citep{P06, Yelda14}, 
	and its apocentre coincides with the inner rim of that disc. Nevertheless, the SPH technique has a tendency for artificially clumping gas 
	\citep{H13}, which raises doubts on how physical the clump formation is in such models. In this context, this work aims to test independently 
	the clump formation as result of colliding stellar winds in the central parsec of the Milky Way.

	Massive stars have significant phases of mass loss through their lives, in which their outflows can be accelerated up to supersonic speeds due to 
	radiation pressure. When two of these stars are at short separations, for example in a binary system, their winds collide and generate a hot slab of 
	shocked gas. Depending on the ability of this gas to cool down, we can identify different regimes. If the shocked gas from both stellar winds 
	cool down rapidly, a cold thin shell will be produced centred at the contact discontinuity (hereafter CD) of the wind-wind interaction zone. This slab 
	will be subject to strong instabilities such as the \textit{Non-linear Thin Shell Instability} \citep[hereafter NTSI,][]{V94}. In the case that only one of the 
	winds is highly radiative, a cold thin shell will also be formed; but if instabilities are excited they will be damped by the thermal pressure of the hot shocked 
	gas from the other wind \citep{V83}. When none of the winds are radiative, the Kelvin-Helmholtz instability (hereafter KHI) is the only instability that can be 
	excited \citep{S92}, as it only requires a velocity difference between the winds. However, KHI can also act on top of thin-shell instabilities for radiatively efficient winds.
	
	State-of-the-art numerical modelling \citep{P09,V10,L11,K14} highlights the high computational cost of realistic 
	simulations of unstable colliding wind systems. High spatial and time resolution, plus many physical ingredients, such as gravity, driving of the winds, radiative cooling and 
	orbital motion, are crucial to build realistic models and track the growth of instabilities. In this context, we take an alternative analytical approach and study the thermal evolution of 
	the hot slab created by the colliding winds. This allows us to predict under which conditions, out of a wide parameter space, the NTSI grows; and to estimate the possible 
	resulting clump masses.
	The paper is divided as follows: Section~\ref{sec:cooling} describes the main cooling diagnostic we use throughout this work and Section~\ref{sec:clumps} presents our 
	model and the results of our parameter space study. In Section~\ref{sec:colliding} we discuss how likely the formation of the G2 cloud is through colliding winds based on our results. 
	Then in Section~\ref{sec:discussion} we discuss the limitations of our results and argue for colliding winds binaries as clump sources in the Galactic Centre. 
	Finally, in Section~\ref{sec:conclusions} we present our conclusions and outlook.
\section{Radiative cooling and thin shells}
	\label{sec:cooling}
	The role of radiative cooling is what determines the thickness, density and temperature of the shocked gas layer. 
	If cooling is efficient, the wind collision will produce a dense, thin layer of cold gas, which can easily 
	be subject to the so-called \textit{thin shell instabilities}. In their numerical models, \cite{S92} identified two such instabilities: 
	the NTSI, later explained analytically by \cite{V94}, and  another damped instability consistent with the earlier description by \cite{V83}. 
	Moreover, \cite{D93} described another thin shell instability: the \textit{Transverse Acceleration Instability} (TAI) that mainly takes place off the two-star axis. 
	\cite{L11}, modelling unstable colliding winds systems numerically, concluded that the NTSI is the instability that dominates 
	the cold slab evolution due to its large scale perturbations.
	
\begin{figure}
		\includegraphics[width=0.525\textwidth]{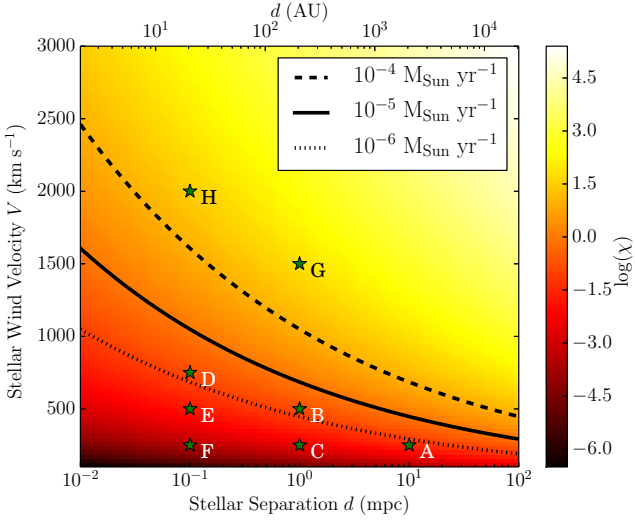}
		\caption{Cooling parameter as colour map on logarithmic scale which was obtained using Equation \ref{eq:chi} and fixing the mass loss 
		rate to 10$^{-5}$ $\msun$ yr$^{-1}$. It is shown as a function of the stellar separation $d$ (for identical stars) on the x-axis and the stellar wind velocity 
		$V$ on the y-axis. The solid line shows $\chi=1$ which separates the two regimes: adiabatic (above, $\chi>1$) and roughly isothermal (below, 
		$\chi<1$) winds. Dashed and dotted lines show the $\chi=1$ boundaries for mass loss rates of 10$^{-4}$ and 10$^{-6}$ $\msun$ yr$^{-1}$, 
		respectively. Green stars labelled A--H indicate the parameters of models we study in more detail in \S~\ref{sec:clumps}. Notice that $1\ \rm mpc \sim206$ AU.}
		\label{fig:chi}
	\end{figure}

	In order to predict whether the wind collision will produce a thin slab we need to check how fast the shocked gas cools down. 
	Following the description presented in \cite{S92}, the radiative efficiency of shocked stellar winds can be described by the \textit{cooling parameter} 
	$\chi$, which is the ratio of the cooling timescale to the dynamical timescale,
	\begin{equation}
		\chi = \frac{t_{\rm cool}}{t_{\rm dyn}},
	\end{equation}
	where $t_{\rm cool}$ and $t_{\rm dyn}$ are the cooling and the dynamical timescales, respectively. They are defined as,
	\begin{equation}
		t_{\rm cool}=\frac{3k_{\rm B}T}{2n\Lambda(T)} \ ; \ t_{\rm dyn}=\frac{d_*}{c_{\rm s}},
	\end{equation}
	where $T$ and $n$ are the temperature and number density of the shocked gas, respectively, $\Lambda(T)$ is the cooling function 
	which we will specify in the following section, $k_{\rm B}$ is the Boltzmann constant, $d_*$ is the distance between the star 
	and the CD and $c_{\rm s}$ is the shocked gas sound speed. Therefore, for $\chi$ lower than unity, the gas can cool down faster 
	through radiative cooling than through adiabatic expansion. On the contrary, $\chi$ larger than one means that radiative cooling is not efficient.
	Following \cite{S92}, but using a cooling function that depends on metal abundance (specified in Section~\ref{sec:an}) 
	and temperature rather than being constant (see Equation~\ref{eq:dis}) we can calculate the cooling parameter for each stellar wind given its properties, 
	\begin{equation}
		\chi \approx \frac{1}{2}\frac{V_8^{5.4}d_{*12}}{\dot{M}_{-7}},
		\label{eq:chi}
	\end{equation} 
	where $V_8$ is the terminal wind speed $V$ in units of 1000~km~s$^{-1}$, $d_{*12}$ is the distance between the star and the CD $d_*$ in units of $10^{12}$~cm 
	and $\dot{M}_{-7}$ is the mass loss rate $\dot{M}$ in units of 10$^{-7}$ $\msun$~yr$^{-1}$. In Figure \ref{fig:chi}, we show $\chi$ as a function 
	of $V$ and the stellar separation of identical stars $d=2d_*$ for $\dot{M}=10^{-5}$ $\msun\,$yr$^{-1}$ fixed, which is the typical value for 
	the Wolf-Rayet stars in the Galactic centre (see \S~\ref{sec:colliding}). We have highlighted the boundary $\chi = 1$ to separate the two cooling regimes. 
	We will concentrate on the $\chi <1$ region, which is where the NTSI excitation is possible.
\section{NTSI and clump formation}
	\label{sec:clumps}
	As discussed in the previous section, if the slab gas can cool down rapidly, different thin shell instabilities can be excited.
 	The dominant instability is the NTSI, which is the result of the misalignment of the thermal pressure within the cold slab 
	(which always acts perpendicular to the slab) and the ram pressure of the wind (always acting parallel to the wind direction).
	This generates a convergent flux onto perturbation knots where gas accumulates, and a subsequent mixing of both phases of the material.
	In this section we use the description of \cite{V94} to calculate the allowed wavelength range of the NTSI which can grow. 
	\subsection{Range of unstable wavelengths}
		The growth of the NTSI can occur only for wavelengths shorter than those coordinated by sound waves: $\lambda < c_{\rm s} t$. 
		On the other hand, the shortest unstable wavelength is given by the slab thickness $l(t)$, i.e., $\lambda \simgt  2l(t)$. 
		Moreover, the NTSI can grow only if the slab perturbation has an initial amplitude at least comparable to the thickness of the layer. 
		Notice that the criteria depend on the time $t$ elapsed since the formation of the slab.
		The thickness of the slab generated by isothermal flows is given by $l(t)=2V_{\rm n}t$, where $V_{\rm n} = V/(\mathcal{M}^2-1)$ is the 
		velocity at which the slab thickens, $\mathcal{M}=V_{\rm s}/c_{\rm s}$ is the Mach number, and $V_{\rm s}=V+V_{\rm n}$ is the shock 
		speed. Notice that for the limit of high Mach number $V_{\rm s}\approx V$, and that the more supersonic the shock, the slower the slab increases its width.
	\subsection{Analytic model}
		\label{sec:an}
		In our model we consider two identical stars that are fixed in space. They have a given 
		mass loss rate and wind terminal velocity, and are separated by a given distance. Respectively, these three quantities $\dot{M}$, $V$ 
		and $d$, are the input parameters of the model. We follow the time evolution of the gas within the slab formed when the winds collide
		by studying its integrated density and thermal evolution, including a radiative cooling term.  We define the relevant densities as,
		\begin{equation}
			\Sigma_{\rm slab}=\rho_{\rm slab}L \label{eq:sys0},
		\end{equation}
		where $\rho_{\rm slab}$ is the volumetric density inside the slab, $\Sigma_{\rm slab}$ is the surface density of the slab and $L$ is the 
		width of the slab.  All these quantities evolve  and we study them as a function of their age, i.e., the time after the wind collision. 
		As we are studying a symmetric 
		system (identical stars with identical winds), we model one side of the system. Therefore, $L(t)$ will be the distance from the CD to the 
		discontinuity of the shocked and the free wind region. As the slab is supported by thermal pressure, in this point the thermal and the wind 
		ram pressure have to be balanced, which is a reasonable assumption considering that the thermal energy is 
		negligible in the free-wind region that is highly supersonic ($\mathcal{M}\geq20$). From this model, we can write the hydrodynamical equations 
		as follows,
		\begin{eqnarray}
			\frac{d}{dt}\Sigma_{\rm slab}	& = & \rho_{\rm wind}V, \label{eq:sys1}\\
			P_{\rm slab}				& = & \rho_{\rm wind}V^2, \label{eq:sys2}\\
			\frac{3k_{\rm B}}{2\mu m_{\rm H}}\frac{d}{dt}\left(\Sigma_{\rm slab}T_{\rm slab}\right) & = & H_{\rm shock} + S_{\rm cool}, \label{eq:sys3}
		\end{eqnarray}
		where $P_{\rm slab}$ is the pressure in the slab, $\mu$ is the mean molecular weight, $m_{\rm H}$ is the proton mass, $H_{\rm shock}$ is the 
		mechanical heating term (i.e. kinetic energy flux from the wind per unit surface), and $S_{\rm cool}$ is the energy dissipation term through 
		radiative cooling. These source terms are given by
		\begin{eqnarray}
			H_{\rm shock}	& = & \frac{1}{2}\rho_{\rm wind}V^3, \label{eq:sys4}\\	
			S_{\rm cool}	& = & \frac{\Sigma_{\rm slab} \rho_{\rm slab}}{\mu^2m_{\rm H}^2}\Lambda(T_{\rm slab}), \label{eq:sys5}
		\end{eqnarray}
		where the cooling function $\Lambda(T)$ is an analytical approximation for optically thin 
		radiative cooling made by \cite{C05} following \cite{S93},
		 \begin{equation}
		 	\Lambda (T) = 6.0\times10^{-23}\left(\frac{Z}{3}\right)\left(\frac{T}{10^7\rm K}\right)^{-0.7} \ {\rm erg \ cm^3 \ s^{-1} },
			\label{eq:dis}
		 \end{equation}
		where $Z$ is the metal abundance relative to solar. We set $Z=3$ and discuss the impact of this choice in Section~\ref{sec:limit}.
		Combining the system of Equations~\ref{eq:sys1},~\ref{eq:sys2},~\ref{eq:sys3},~\ref{eq:sys4},~\ref{eq:sys5}; and assuming an ideal gas 
		$P_{\rm slab}=\rho_{\rm slab}k_{\rm B}T_{\rm slab}/(\mu m_{\rm H})$, we can derive a single differential equation that describes the thermal evolution 
		of the slab:
		\begin{equation}
			\frac{3k_{\rm B}}{2\mu m_{\rm H}}\frac{d}{dt}T_{\rm slab}=\frac{1}{2}\frac{V^2}{t}-\frac{\rho_{\rm wind}V^2}{k_{\rm B}\mu m_{\rm H}}\frac{\Lambda(T_{\rm slab})}{T_{\rm slab}}-\frac{3k_{\rm B}}{2\mu m_{\rm H}}\frac{T_{\rm slab}}{t}.
		\end{equation}
		The first term on the right hand side represents the mechanical heating of the shock; the second, the radiative cooling; and the third, the work 
		done by the slab. Furthermore, the slab density can be easily computed making use of Equation~\ref{eq:sys2} and the ideal gas equation of state. 
		Thus, the slab width time evolution can be calculated using Equations~\ref{eq:sys0} and~\ref{eq:sys1} to obtain
		\begin{equation}
			\frac{d}{dt}\left[\rho_{\rm slab}(t)L(t)\right]=\rho_{\rm wind}V \Rightarrow  L(t)=\frac{\rho_{\rm wind}}{\rho_{\rm slab}(t)}Vt.
		\end{equation}
		From this expression we note that if the slab does not cool down (i.e. the slab density remains roughly constant), the slab will increase its width linearly with 
		time; on the contrary, if the slab gets denser very rapidly it may overcome the linearity dependence with time and collapse as a thin shell which grows 
		very slowly due to the fact that $\rho_{\rm wind}/\rho_{\rm slab}\ll1$. 
		\begin{table}
			\centering
			\caption{Parameters of each model presented in this work. Column 1: Model name. Column 2: wind terminal velocity. 
				Columns 3 and 4: stellar separation in mpc and AU, respectively. Column 5: cooling parameter $\chi$ estimated from Equation~\ref{eq:chi} 
				and $\dot{M}$ fixed to 10$^{-5}$ $\msun$ yr$^{-1}$.}
			\label{tab:models}
			\begin{tabular}{@{}crrrr@{}}
				\hline
				Model	&	$V$			&	\multicolumn{2}{c}{$d$}	&	\multicolumn{1}{c}{$\chi$}\\
						&	(km s$^{-1}$)	&	(mpc)	&\multicolumn{1}{c}{(AU)}	&\\
				\hline
				\hline
				A		&	250			&	10.0		&	$2.06\times10^3$	&	0.0433\\
				B 		&	500			& 	1.0		&	$2.06\times10^2$	&	0.1827\\
				C		&	250			&	1.0		&	$2.06\times10^2$	&	0.0043\\
				D		&	750			&	0.1		&	$2.06\times10^1$	&	0.1632\\
				E		&	500			&	0.1		&	$2.06\times10^1$	&	0.0183\\
				F		&	250			&	0.1		&	$2.06\times10^1$	&	0.0004\\
				G		&	1500			&	1.0		&	$2.06\times10^2$	&	69.89\\
				H		&	2000			&	0.1		&	$2.06\times10^1$	&	32.57\\
				\hline	
			\end{tabular}
		\end{table}
		With this analysis we can calculate the NTSI growth criteria as a function of the age of the slab (or time after the wind collision). 
		Although the model is simple, it can provide us with information about the instabilities that can take place in order to estimate 
		sizes and masses of the possible clumps. It is important to mention that throughout this analysis we are not attempting to follow 
		the evolution of the instability, but to study under which conditions it can be triggered.
	\subsection{Parameter study}
		Using the previously described procedure, we explored the parameter space shown in Figure~\ref{fig:chi}. 
		We first present in detail a few relevant and representative examples (see Table~\ref{tab:models}). In all these models we 
		kept $\dot{M}=10^{-5}$ $\msun$ yr$^{-1}$ fixed and stopped the calculation after the gas has cooled down significantly, reaching a 
		temperature of 10$^4$ K.  This is a reasonable temperature floor because of the presence of many hot, young stars emitting ionising 
		radiation that prevents the temperature to drop to lower values\footnote{We also tested stopping the cooling at 10$^5$ K, finding that density 
		values are systematically smaller by one order of magnitude.}.
		\begin{figure*}
			\includegraphics[width=0.475\textwidth]{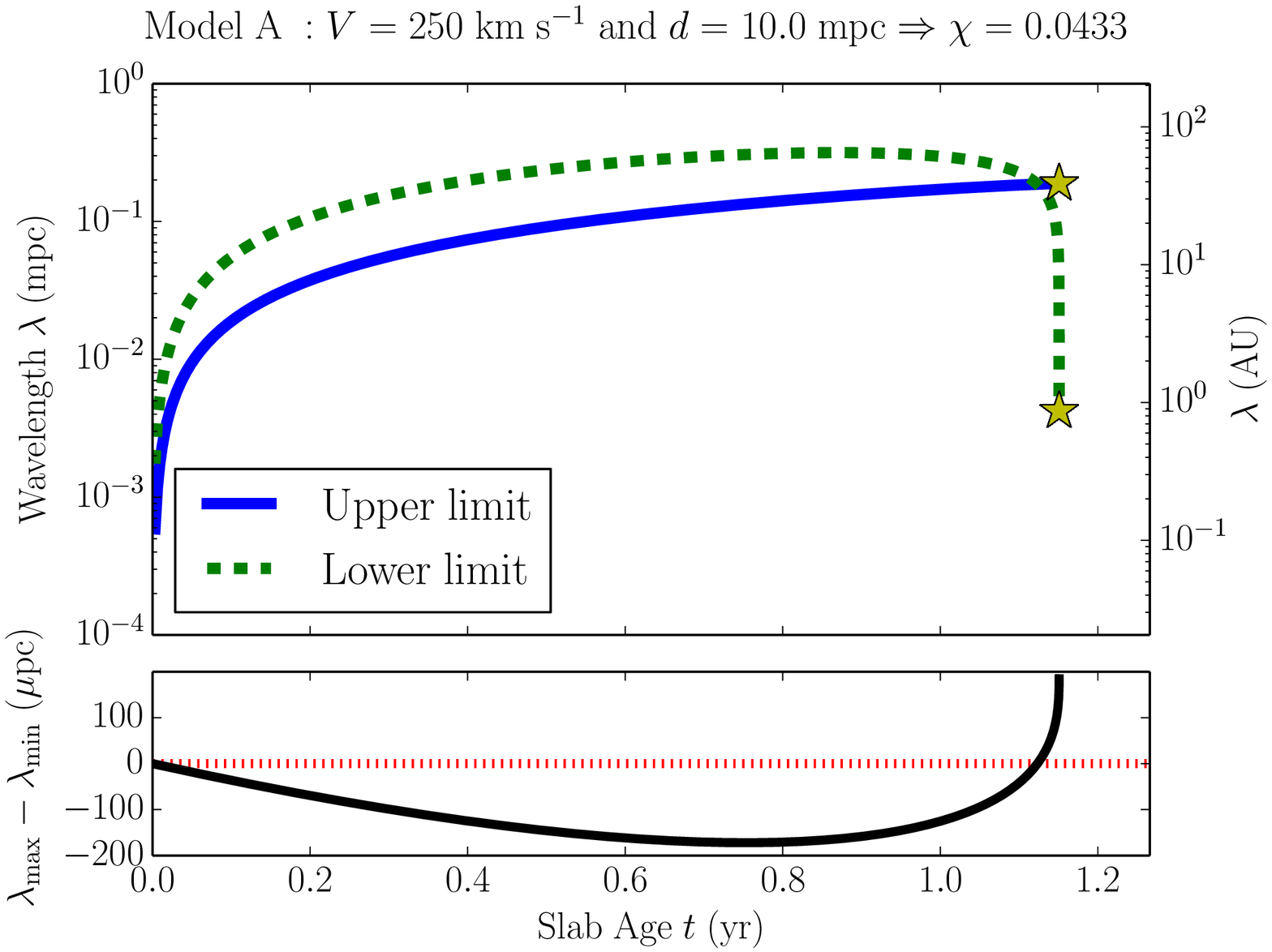}
			\includegraphics[width=0.475\textwidth]{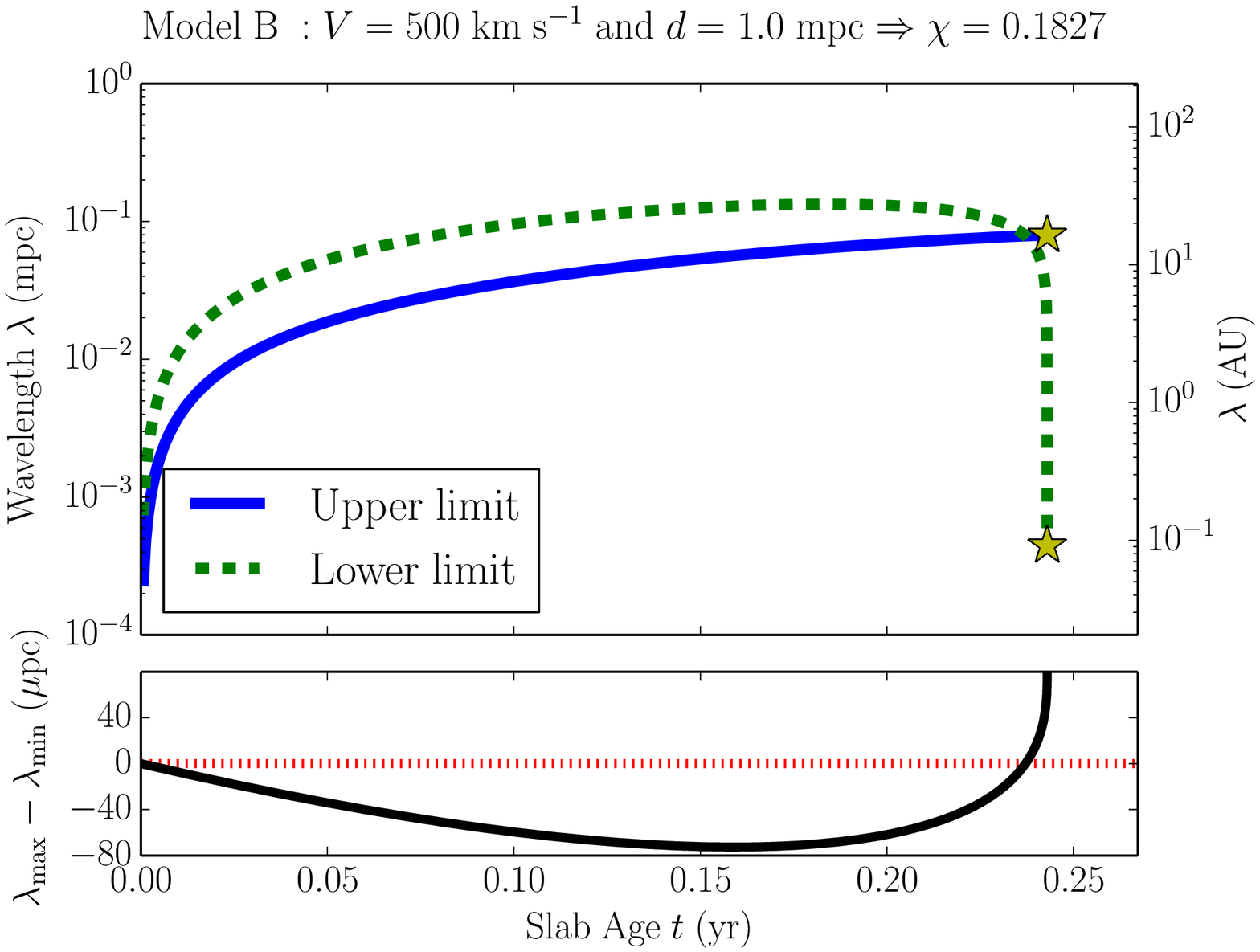}
			\includegraphics[width=0.475\textwidth]{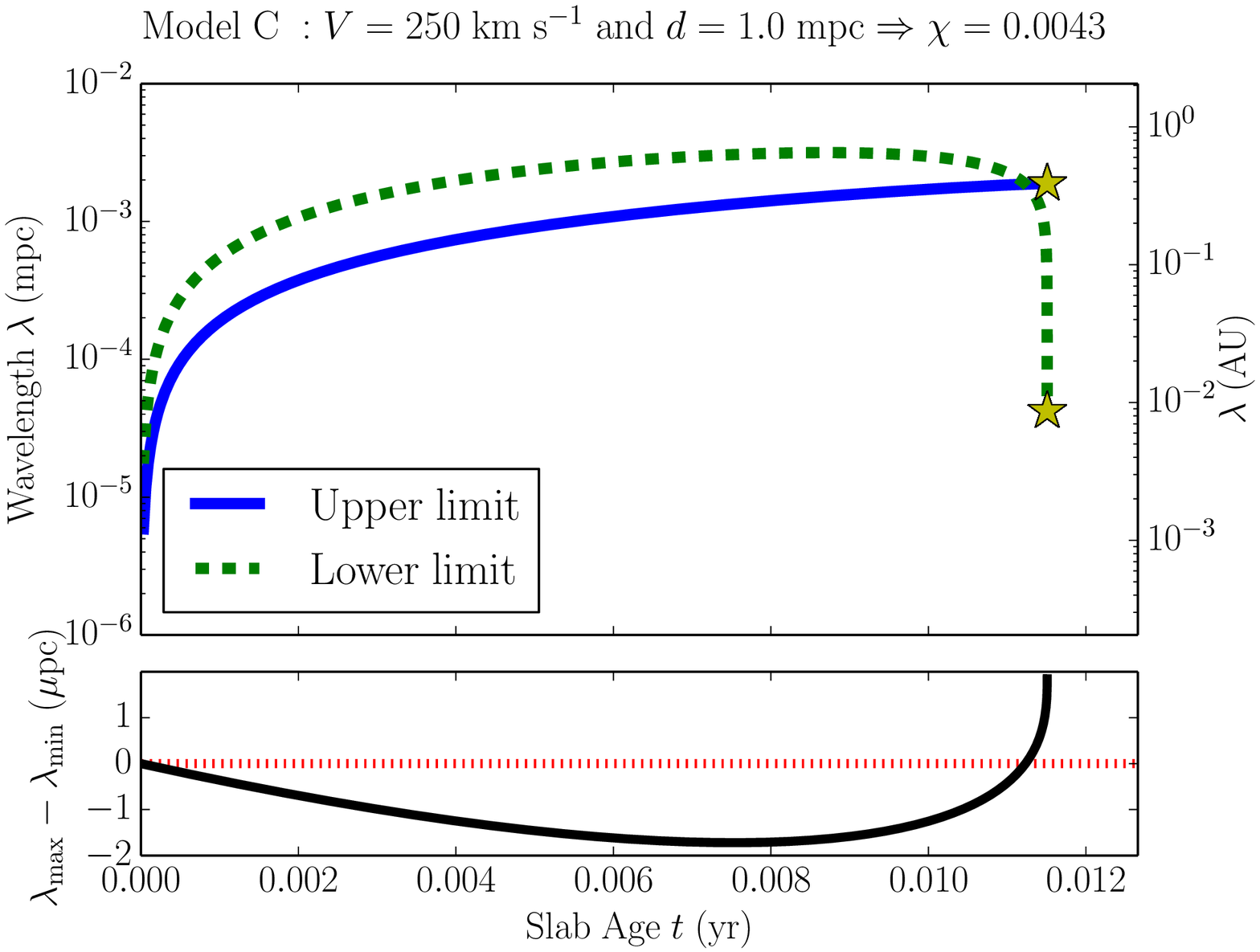}
			\includegraphics[width=0.475\textwidth]{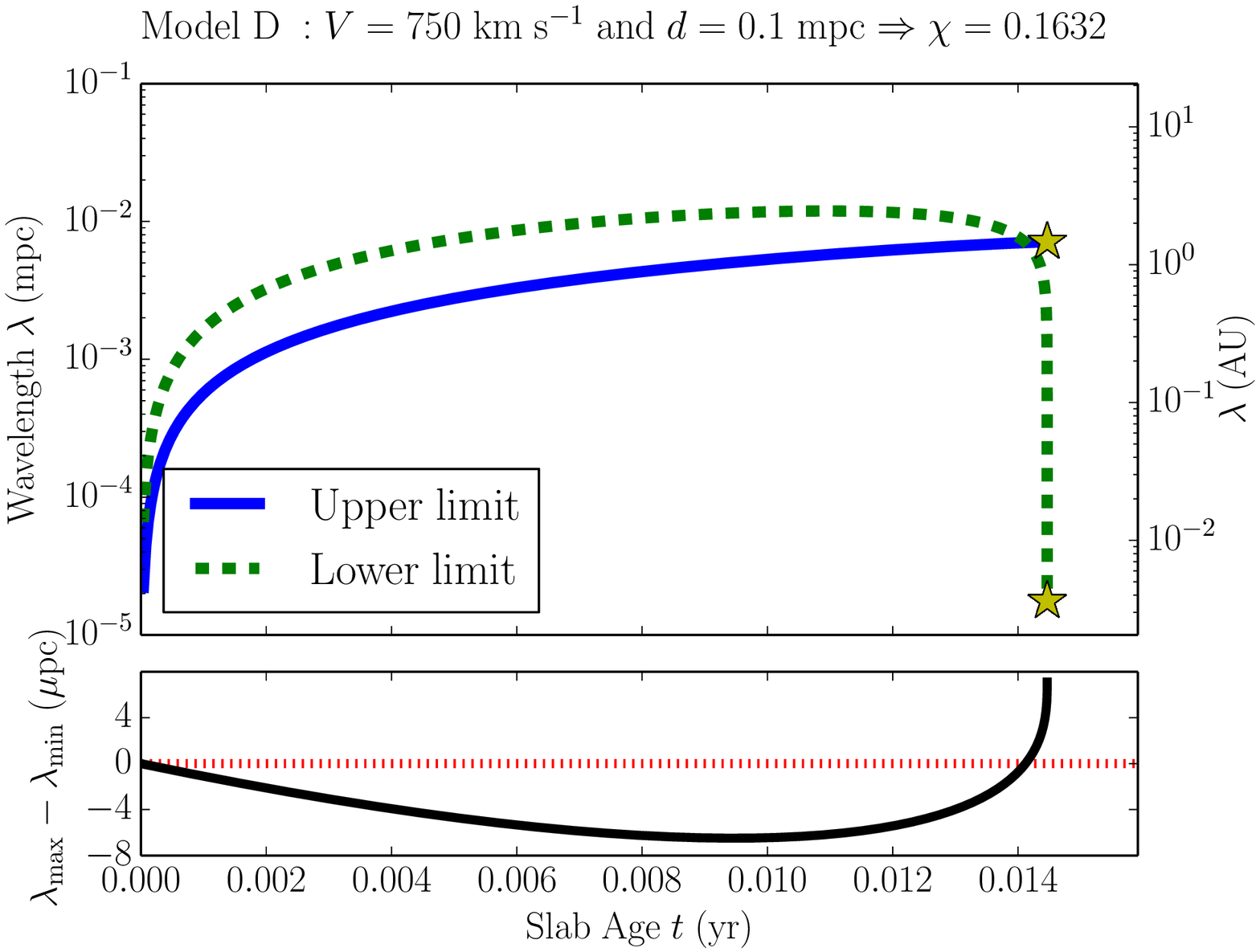}
			\includegraphics[width=0.475\textwidth]{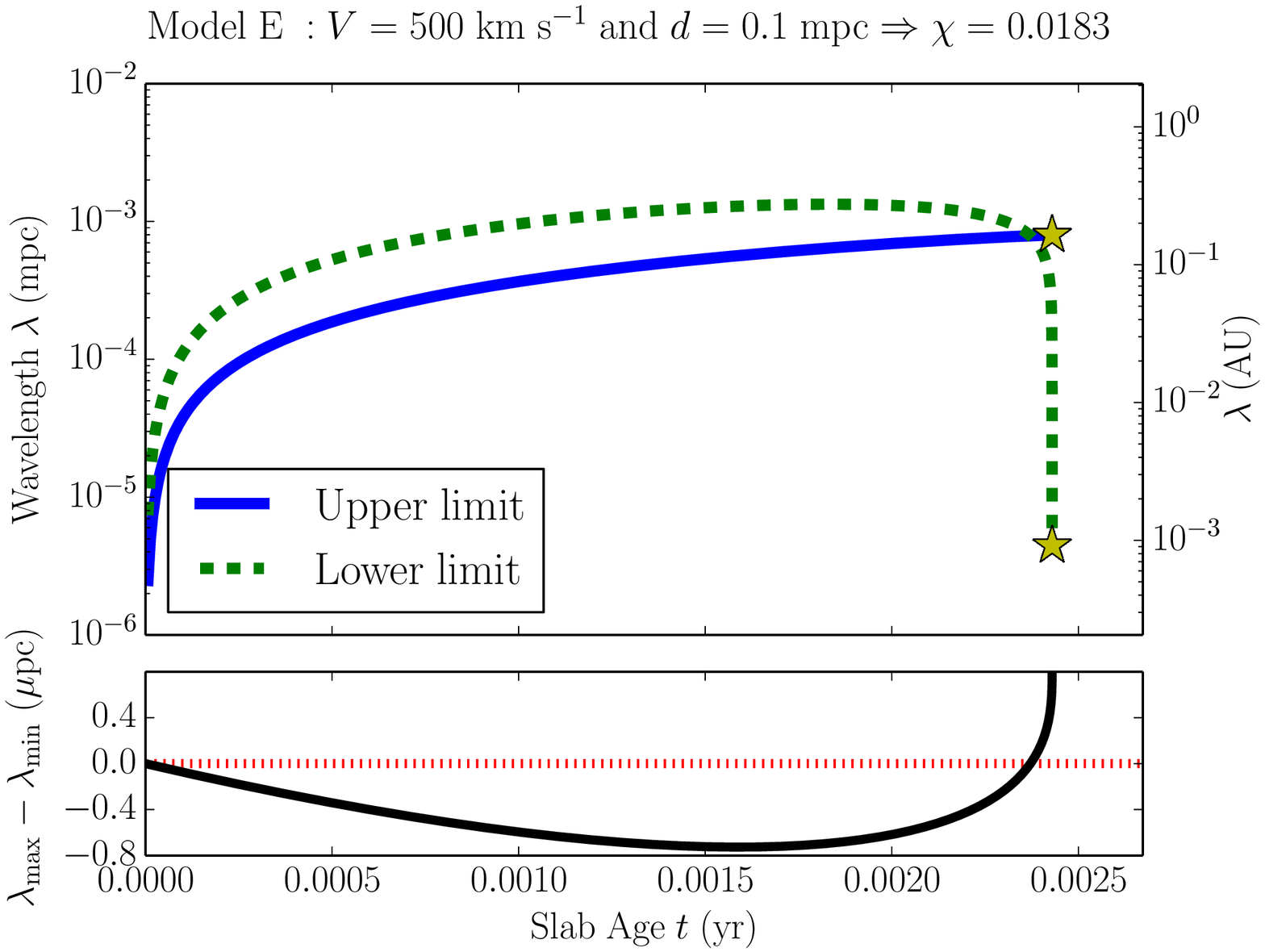}
			\includegraphics[width=0.475\textwidth]{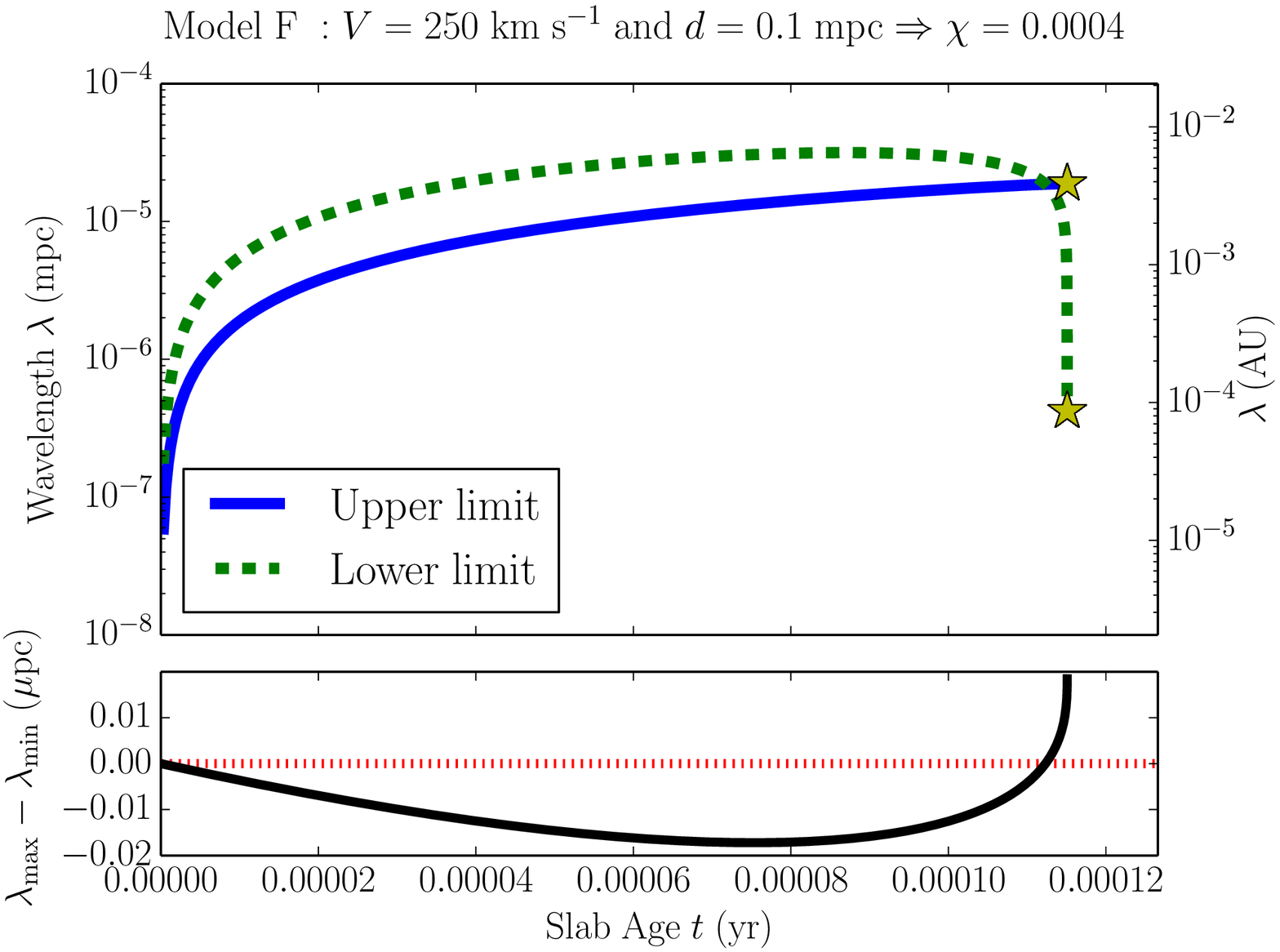}
			\caption{Evolution of the NTSI wavelength criteria for the radiative models ($\chi<1$).
			 On the upper panels we show the time evolution of the upper and lower limits of unstable wavelengths for the NTSI obtained 
			from our semi-analytic prescription. Instabilities can grow only at the end of each time evolution, 
			when a range of allowed wavelengths exists. Yellow stars represent the final state of the system when the gas has reached 10$^4$ K. 
			On the lower panels, the thick black line represents the difference between the upper and lower limits, 
			$\lambda_{max}-\lambda_{min}$, and the dotted red line is fixed at zero for reference. 
			Note the  different scales in the x- and y-axis between models.}
			\label{fig:crit}	
		\end{figure*}
		\begin{figure*}
			\includegraphics[width=0.475\textwidth]{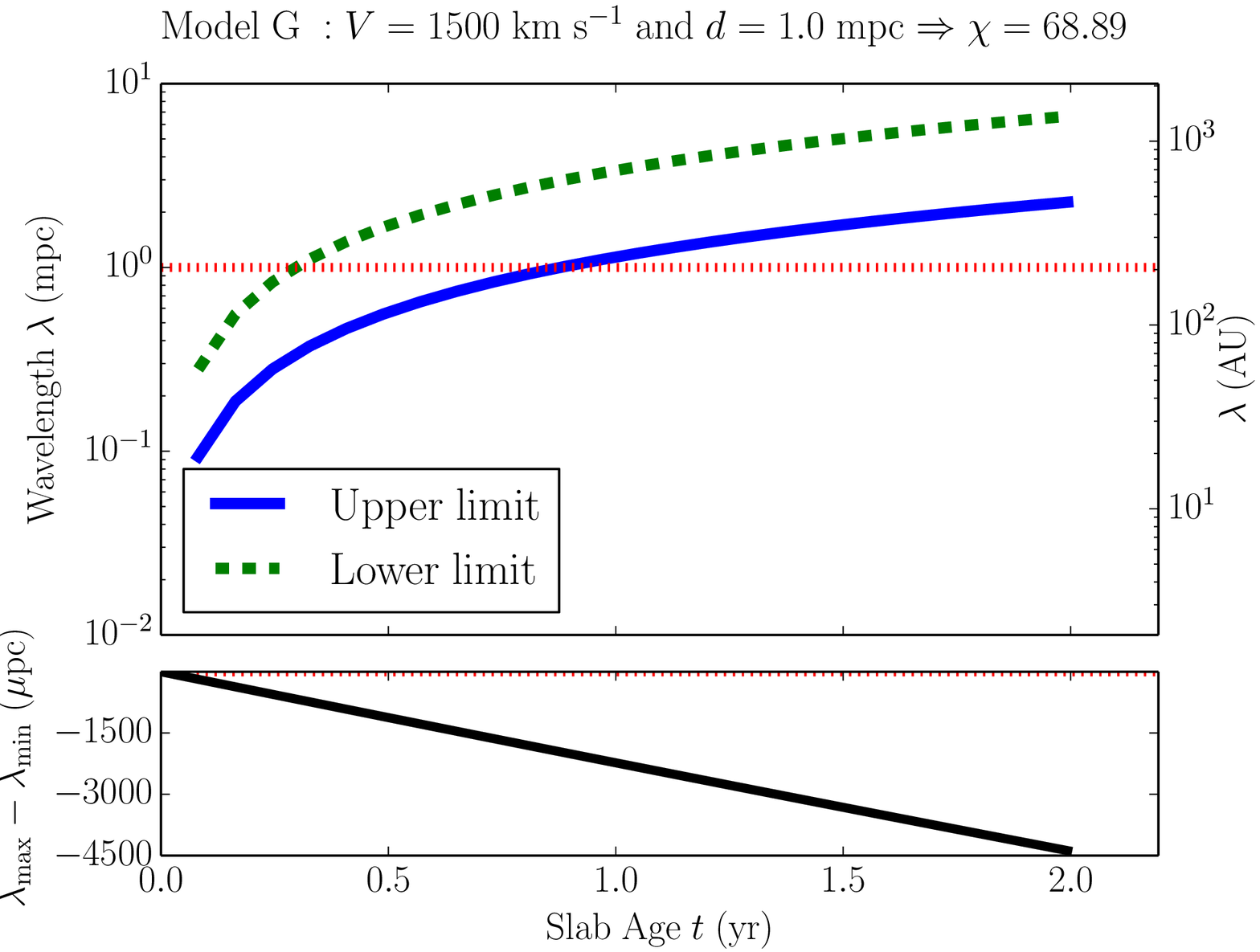}
			\includegraphics[width=0.475\textwidth]{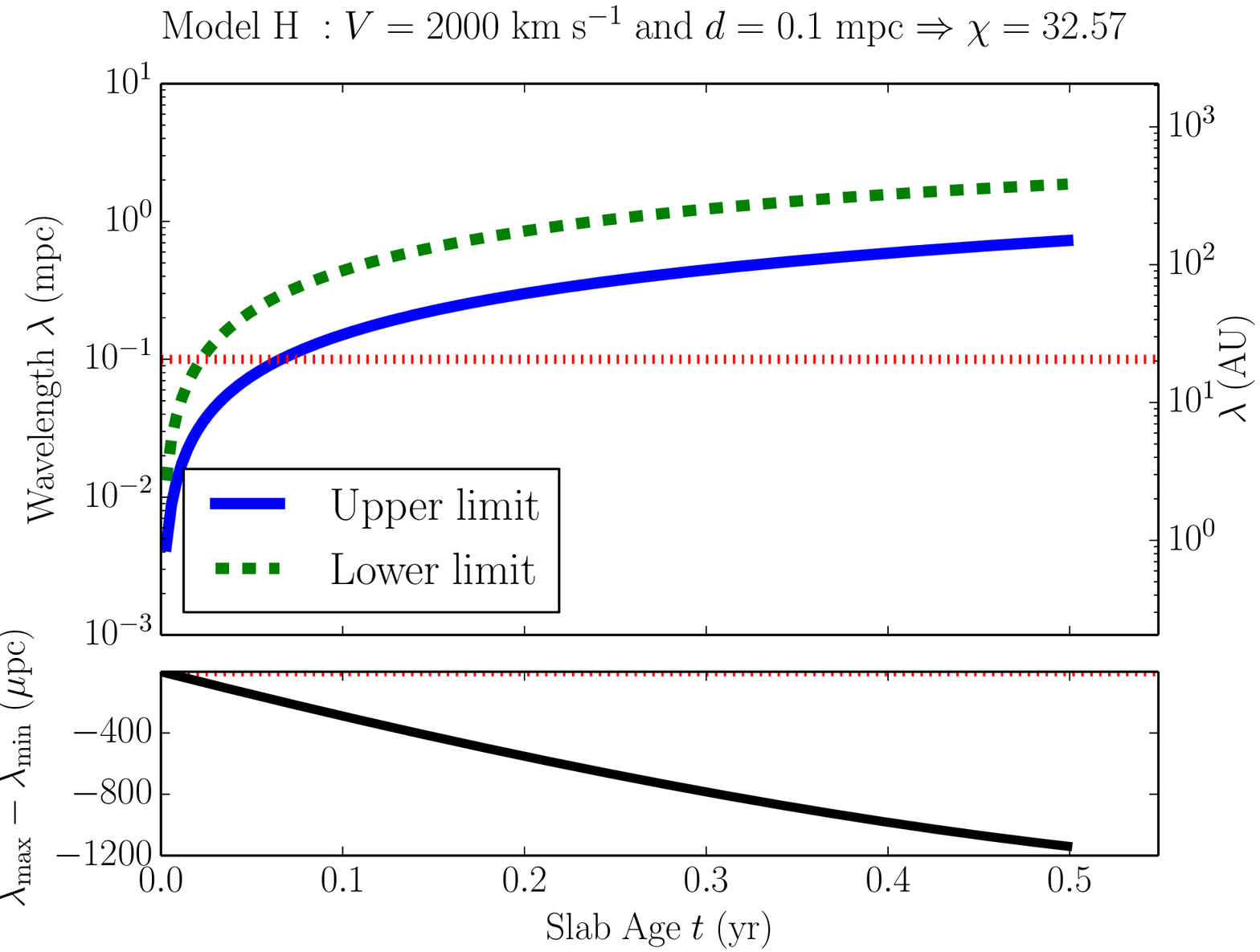}
			\caption{Same as Figure~\ref{fig:crit} but for the adiabatic models ($\chi>1$) G and H. At some point the models become unphysical,
			as the slab width is comparable to the stellar separation (red dotted line).  We do not expect NTSI excitation for these models.}
			\label{fig:crit-ad}	
		\end{figure*}
		\begin{table*}
			\centering
			\caption{Results of our estimates for each model. Column 1: model name. Columns 2 and 3: unstable wavelength range obtained from our analysis in mpc and AU, respectively. 
			Column 4: slab density obtained estimated from our calculations. Column 5: clump mass range computed assuming spherical clumps and uniform density 
			with radius equal to the instability wavelengths. Column 6: growth timescale estimated assuming an initial amplitude equal to the instability wavelength and 
			the slab sound speed at the temperature floor ($10^4$ K).}
			\begin{tabular}{@{}cccccc@{}}
				\hline
				Model	&	\multicolumn{2}{c}{Unstable $\lambda$}					&	$\rho_{\rm slab}$	&	Clump Mass Range			&	Growth Timescale	\\	
						&	$\rm (mpc)$				&	$\rm (AU)$			&	$(\rm g\ cm^{-3})$	&	$(\mearth)$				&		$\rm (yr)$		\\
				\hline
				\hline
				A		&	$(4-190)\times10^{-3}$		&	$(8-390)\times10^{-1}$	&	$2.4\times10^{-18}$	&	$3.6\times10^{-6}-3.3\times10^{-1}$		&$(3-120)\times10^{-1}$		\\
				B		&	$(5-800)\times10^{-4}$		&	$(1-160)\times10^{-1}$	&	$4.7\times10^{-16}$	&	$8.8\times10^{-7}-4.9\times10^{0}$		&$(3-50)\times10^{-1}$		\\
				C		&	$(4-190)\times10^{-5}$		&	$(8-390)\times10^{-3}$	&	$2.4\times10^{-16}$	&	$3.6\times10^{-10}-3.3\times10^{-5}$	&$(3-120)\times10^{-3}$		\\
				D		&	$(2-710)\times10^{-5}$		&	$(4-1460)\times10^{-3}$	&	$7.1\times10^{-14}$	&	$8.5\times10^{-9}-5.3\times10^{-1}$		&$(1-470)\times10^{-3}$		\\
				E		&	$(5-800)\times10^{-6}$		&	$(1-160)\times10^{-3}$	&	$4.7\times10^{-14}$	&	$8.8\times10^{-11}-4.7\times10^{-4}$		&$(3-50)\times10^{-4}$		\\
				F		&	$(4-190)\times10^{-7}$		&	$(8-390)\times10^{-5}$	&	$2.4\times10^{-14}$	&	$3.6\times10^{-14}-3.3\times10^{-9}$	&$(3-120)\times10^{-5}$	\\
				G		&					-		&	-					&	$2.6\times10^{-19}$	&				-					&	-\\
				H		&					-		&	-					&	$2.6\times10^{-21}$	&				-					&	-\\
				\hline
			\end{tabular}
			\label{tab:sim}
		\end{table*}
		
		The time evolution of the unstable wavelength criteria for the NTSI are shown for all models in Figures~\ref{fig:crit} and~\ref{fig:crit-ad}. 
		Figure~\ref{fig:crit} shows the models with $\chi<1$. Here, for most of the evolution of the slab, the upper limit (solid blue line) 
		is {\it below} the lower limit (dashed green line), meaning that the instability cannot grow as there is no unstable wavelength range.  
		Only at the very end of each calculation the upper limit is above the lower limit and therefore the instability develops. The sudden 
		drop in the lower limit is due to the fast slab gas compression once cooling becomes much more efficient as the temperature decreases.
		In all these cases we registered the unstable length range and the final density values in Table~\ref{tab:sim}. 
		It is important to remark that for all these cases we remain in the thin shell regime, i.e., $L/d\ll1$. 
		Therefore we do not expect radiative cooling to be faster than the thermal response of the slab, making our assumption of pressure 
		equilibrium at the shock valid.
		
		Assuming spherical symmetry, we used the unstable wavelength range to compute a range of masses for clumps (see Table~\ref{tab:sim}). 
		However, these values must be interpreted only as upper limits, because we have assumed the clump radii to be about the size of the unstable wavelengths, 
		which is probably an extreme case.
		From the results we see that larger stellar separations will result in larger clumps.  However, these clumps are less dense due to the winds being significantly diluted before they collide. 
		On the contrary, smaller separations produce smaller and denser clumps. The combination of these two factors sets the clump masses in a non-trivial manner.
		Our results show that~\textit{Model A},~\textit{Model B} and~\textit{Model D} could generate clumps with masses $>0.1 \mearth$ and the most 
		massive ones would be generated by \textit{Model B}, reaching G2-like masses ($\sim5 \mearth$).\footnote{G2 could also have formed 
		by mergers of smaller clumps, but we cannot address this option with our current approach.} 

		On the other hand, Figure~\ref{fig:crit-ad} shows the models with $\chi>1$, in which we see that  the width of the slab (dashed green line) 
		becomes larger than the stellar separation (dotted red line) before there is an unstable wavelength range. Thus, we deem these models not 
		physical, as our treatment cannot describe these systems properly, and we expect no clump formation through NTSI for this parameter range.

\begin{figure}
			\includegraphics[width=0.51\textwidth]{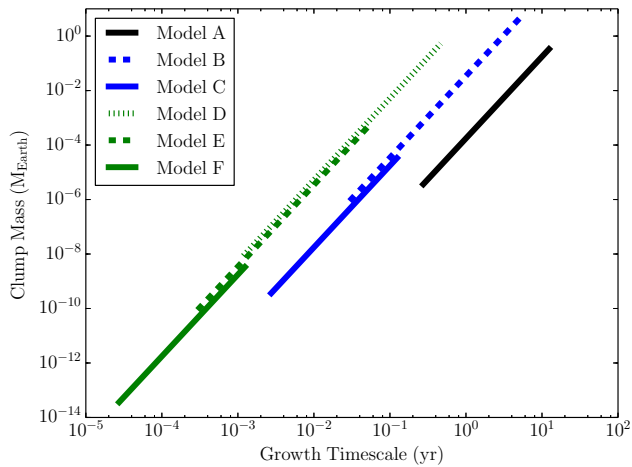}
			\caption{Clump masses (assuming spherical symmetry) generated through the NTSI, 
			as a function of the instability growth timescale (assuming an initial amplitude comparable to the instability wavelength). 
			Different stellar separations are shown with different colours: green, blue and black stand for 0.1, 1 and 10 mpc (20.6, 206 and 2060 AU), respectively,
			while different wind velocities are shown with different line styles:  solid, dashed and dotted represent 250, 500 and 750 km s$^{-1}$.}
			\label{fig:clumps}	
		\end{figure}

		Although these results show that clump formation might occur for the $\chi<1$ models, we need to check on which timescales this process takes place. 
		\cite{V94} showed that the NTSI growth timescale $\tau$ is given by
		\begin{equation}
			\tau = \frac{\lambda^{1.5}}{\zeta^{0.5} c_{\rm s}},
			\label{eq:tau}
		\end{equation}
		where $\zeta$ is the initial amplitude of the perturbation. If we assume $\zeta\sim\lambda$, the growth timescale is simply given 
		by the sound crossing timescale, i.e., $\tau\sim\lambda/c_{\rm s}$. Estimations of $\tau$ under this assumption for each of our 
		models are presented in Table~\ref{tab:sim}. Moreover, we present the clump masses as a function of the growth timescale for 
		each model in Figure~\ref{fig:clumps}, where we see that models with shorter stellar separations ($d=0.1$ mpc in green lines) 
		tend to be the ones that can create clumps the quickest. This is due to these winds being less diluted when they collide, making the cooling 
		more efficient. We can compensate this effect by increasing the wind speed, as a hotter slab would take longer to radiate most of its energy away. 
		For example, comparing cases C and E (solid blue and dashed green, respectively) we see that the combination of different stellar 
		separations and different wind speeds produces clump formation on roughly the same timescale.
		
\begin{figure*}
			\includegraphics[width=0.85\textwidth]{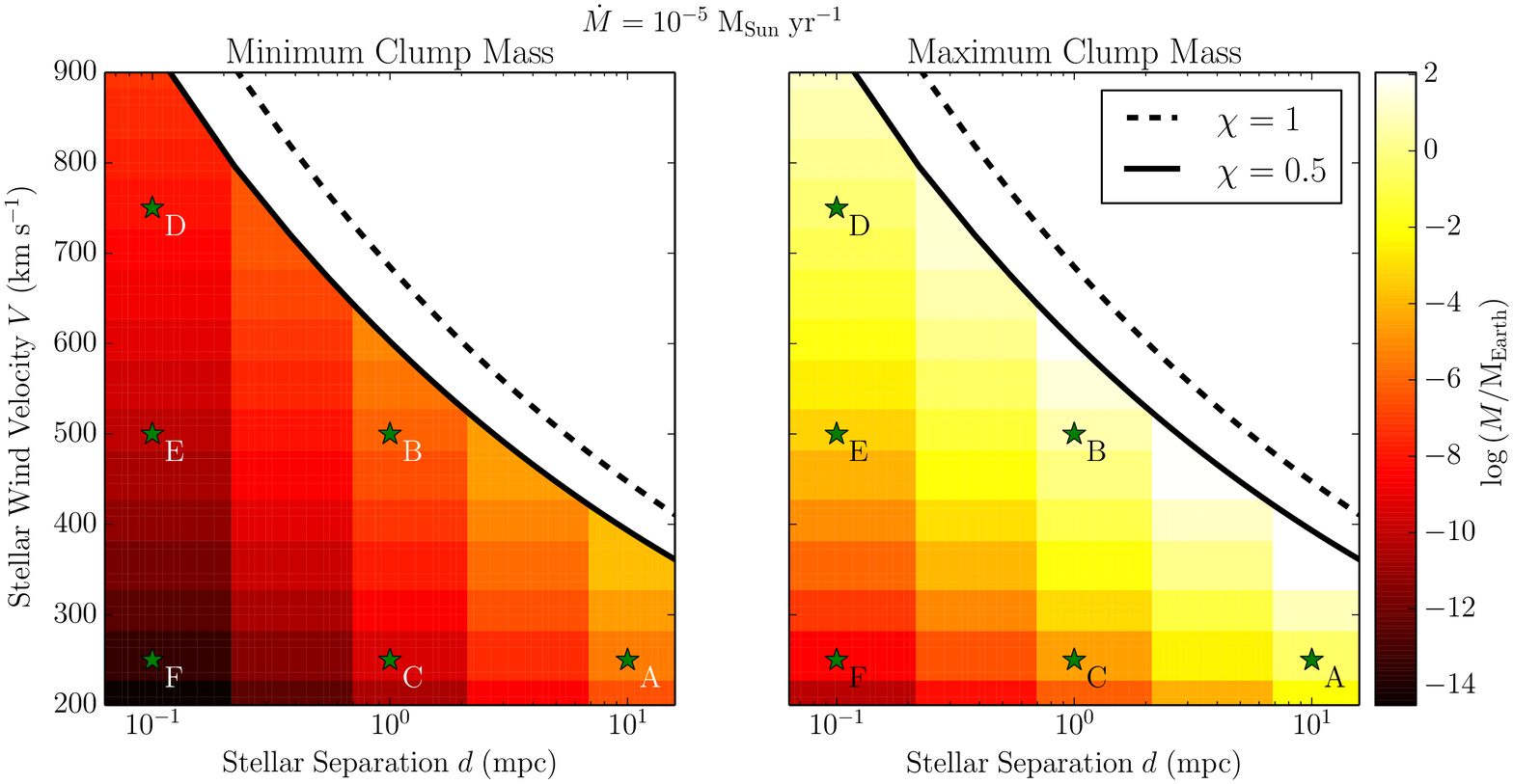}
			\caption{Parameter space $(d,V)$ with $\dot{M}=10^{-5}\ \msun\,$yr$^{-1}$ fixed, displaying clump masses (colours) 
			formed through NTSI obtained with our model. The left and right panels show the minimum (lower limit) and maximum 
			(upper limit) clump mass for a given combination of parameters, respectively. Green star symbols show the 
			models we previously studied in detail. The solid and dashed black lines stand for $\chi=0.5$ and $\chi=1$, respectively.}
			\label{fig:mass-range}
		\end{figure*}

		An interesting point is that the models that can create the most massive clumps have significantly different parameters. 
		This can be explained by the fact that clump masses are proportional to $\lambda^3$, thus the most massive clump possibly generated 
		would have a mass $M\propto\lambda_{\rm max}^3$. Moreover, the upper limit of the unstable wavelength range of the NTSI is given by 
		the sound crossing distance, i.e., $\int c_{\rm s}dt$ integrated over the age of the slab when it reaches $10^4$~K. This tells us that how 
		massive clumps can be, depends on how long it takes for the slab to cool down and become unstable. Then, either high velocity winds 
		and/or larger separations would produce more massive clumps. In order to illustrate this explicitly we have explored our parameter space 
		more extensively, modelling 45 systems in total. We show this in Figure~\ref{fig:mass-range} as a function of the input parameters of our 
		model. The left and right panels present the lower and upper limits of the clump mass range, respectively. Here it is easy to see that for 
		larger separations and/or higher wind velocities the minimum and maximum clump masses are larger. 
		
		We repeated the previous analysis using different mass loss rate values, $10^{-6}$ and $10^{-4}$ $\msun$ yr$^{-1}$, finding that for lower $\dot{M}$ 
		values, clumps can be more massive for the same combination of $(d,V)$, as a less dense slab makes the cooling less efficient, but we 
		would need shorter distances and/or slower winds in order to get radiative winds in the first place, as the $\chi=1$ line moves down in the 
		parameter space (see Figure~\ref{fig:chi}). On the contrary, increasing $\dot{M}$ we obtain less massive clumps for the same $(d,V)$, 
		as a denser slab makes the cooling more efficient, but this would allow larger separations and/or faster winds to generate radiative winds, 
		as the $\chi=1$ line moves up in the parameter space.
		
		To study which clumps will actually form, we take into account Equation~\ref{eq:tau} that shows that shorter wavelengths grow faster.
 		Therefore for any model we would  expect first the formation of the lightest possible clumps. On longer timescales, larger wavelengths 
		would also act, accumulating the small clumps and possibly  merging them to create more massive clumps. To actually predict a clump 
		mass distribution, we require numerical simulations. We defer that study to a forthcoming work.

		In general, we find that for parameters closer to $\chi=1$ (black dashed line in Figure~\ref{fig:mass-range}) we would expect more massive 
		clumps to be formed. However we do not know exactly at what point our approach becomes unphysical as we go closer to the adiabatic 
		regime, as seen in models G and H (see Figure~\ref{fig:crit-ad}). Thus, it is more sensible to explore the parameter space close to $\chi=1$ 
		with numerical simulations. This is why we explored values up to $\chi=0.5$ only (black solid line). Despite this, we see that clump masses span a 
		very wide range of masses and that the creation of clumps as massive as G2 is possible. 
\section{Colliding stellar winds in the Galactic Centre}
	\label{sec:colliding}
	We have studied the parameter space for close pairs of mass-losing stars, checking whether we expect them to form 
	clumps through the NTSI, plus estimating the allowed range of sizes and masses for those clumps. Now, we apply this 
	model to the stars in the central parsec of the Milky Way, as the NTSI could be excited for the colliding winds of 
	massive stars, and explain the origin of the G2 cloud. Our model assumes that the stars are stationary and that the pairs are identical.  
	The first assumption is justified as the time-scales involved in the instability ($\simlt 10\,$yr, e.g., Figure~\ref{fig:clumps}) are 
	typically much shorter than the duration of close stellar encounters of mass-losing stars in the Galactic Centre (see Table~\ref{tab:encon}). 
	The second issue is partially addressed in Section~\ref{sec:asymm}. 
	Our sample of mass-losing stars in the inner parsec of the Galaxy is the same used in the numerical models by 
	\cite{C08,C15} and it is listed in Table \ref{tab:stars}.
	For most of the stars, the wind properties were taken from \cite{M07}, who fitted stellar atmosphere models to spectra obtained with 
	SINFONI at the VLT. For the rest, the wind properties were assigned by \cite{C08} based on their similarity to other stars whose spectra 
	were properly modelled by \cite{M07}. The typical uncertainty in the parameters derived by \cite{M07} is $\sim100$~km s$^{-1}$ for the 
	velocities and  $\sim0.2$~dex for the mass-loss rates, small enough not to influence our conclusions.
		\begin{table}
			\centering
			\caption{Galactic Centre mass-losing star sample taken from \protect\cite{M07} and \protect\cite{C08}. 
			Column 1: star ID. Column 2: star name. Both from \protect\cite{P06}. 
			Column 3:  stellar wind terminal velocity. 
			Column 4: stellar mass loss rate. 
			Columns 5 and 6: twice the radiative-to-adiabatic wind transition distance computed from Equation~\ref{eq:trans} in mpc and AU, respectively.}
  			\begin{tabular}{@{}clrccrr@{}}
  				\hline
				ID	&	Name	&	$V$			&	$\dot{M}$				&\multicolumn{2}{c}{2$d_*^{\rm cool}$}\\
					&			&	(km s$^{-1}$)	&	($\msun$ yr$^{-1}$)		&	(mpc)	&	(AU)\\
				\hline
				\hline
				19	&	16NW	&	600			&	$1.12\times10^{-5}$		&	2.29		&	472\\
				20	&	16C		&	650			&	$2.24\times10^{-5}$		&	2.97		&	612\\	
				23	&	16SW	&	600			&	$1.12\times10^{-5}$		&	2.29		&	472\\
				31	&	29N		&	1000			&	$1.13\times10^{-5}$		&	0.15		&	31\\
				32	&	16SE1	&	1000			&	$1.13\times10^{-5}$		&	0.15		&	31\\	
				35	&	29NE1	&	1000			&	$1.13\times10^{-5}$		&	0.15		&	31\\
				39	&	16NE	&	650			&	$2.24\times10^{-5}$		&	2.97		&	612\\
				40	&	16SE2	&	2500			&	$7.08\times10^{-5}$		&	0.01		&	2\\
				41	&	33E		&	450			&	$1.58\times10^{-5}$		&	15.28	&	3148\\
				48	&	13E4		&	2200			&	$5.01\times10^{-5}$		&	0.01		&	2\\
				51	&	13E2		&	750			&	$4.47\times10^{-5}$		&	2.74		&	564\\
				56	&	34W		&	650			&	$1.32\times10^{-5}$		&	1.75		&	361\\
				59	&	7SE		&	1000			&	$1.26\times10^{-5}$		&	0.16		&	33\\
				60	&	-		&	750			&	$5.01\times10^{-6}$		&	0.31		&	64\\
				61	&	34NW	&	750			&	$5.01\times10^{-6}$		&	0.31		&	64\\
				65	&	9W		&	1100			&	$4.47\times10^{-5}$		&	0.35		&	72\\
				66	&	7SW		&	900			&	$2.00\times10^{-5}$		&	0.46		&	95\\
				68	&	7W		&	1000			&	$1.00\times10^{-5}$		&	0.13		&	27\\
				70	&	7E2		&	900			&	$1.58\times10^{-5}$		&	0.36		&	74\\
				71	&	-		&	1000			&	$1.13\times10^{-5}$		&	0.15		&	31\\
				72	&	-		&	1000			&	$1.13\times10^{-5}$		&	0.15		&	31\\
				74	&	AFNW	&	800			&	$3.16\times10^{-5}$		&	1.37		&	282\\
				76	&	9SW		&	1000			&	$1.13\times10^{-5}$		&	0.15		&	31\\
				78	&	B1		&	1000			&	$1.13\times10^{-5}$		&	0.15		&	31\\
				79	&	AF		&	700			&	$1.78\times10^{-5}$		&	1.58		&	325\\
				80	&	9SE		&	1000			&	$1.13\times10^{-5}$		&	0.15		&	31\\
				81	&	AFNWNW	&	1800			&	$1.12\times10^{-4}$		&	0.06		&	12\\
				82	&	Blum		&	1000			&	$1.13\times10^{-5}$		&	0.15		&	31\\
				83	&	15SW	&	900			&	$1.58\times10^{-5}$		&	0.36		&	74\\
				88	&	15NE	&	800			&	$2.00\times10^{-5}$		&	0.87		&	179\\ 
				\hline
			\end{tabular}
			\label{tab:stars}
		\end{table}
	\subsection{Radiative cooling diagnostic}
	\label{sec:diagn}
\begin{figure}
		\centering
			\includegraphics[width=0.49\textwidth]{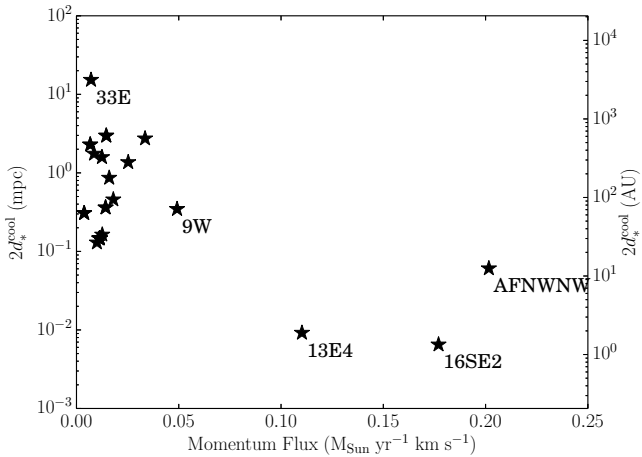}
			\caption{Critical stellar separation in order for
			 their winds to be efficiently radiative (i.e. their collision could produce a \textit{thin shell}) as a function of their wind 
			momentum flux  ($\dot{M}\times V$). The sample plotted corresponds to the one shown in Table \ref{tab:stars}.  Stars with large 
			momentum fluxes are labelled, as they can be important in asymmetric encounters  ``forcing" weaker winds to 
			radiate their energy rapidly, resulting in radiative shocks.  Star 33E is also labelled as it has the largest value of $d^{\rm cool}_*$.}
			\label{fig:diag}
		\end{figure}

		We first estimate  the critical separation between stars in order for their winds to be radiatively efficient. 
		We use the previously defined \textit{cooling parameter} $\chi$, and compute the radiative-to-adiabatic 
		transition separation equating $\chi = 1$,
		\begin{equation}
			d_*^{\rm cool} = 2\times10^{12} {\rm cm}  \, \frac{\dot{M}_{-7}}{v_8^{5.4}} .
			\label{eq:trans}
		\end{equation}
		Pairs of identical stars with separations $d > 2d_*^{\rm cool}$ will produce adiabatic shocks, while 
		those with $d < 2d_*^{\rm cool}$ will result in radiatively-cooled shocks. The value of $2d_*^{\rm cool}$ 
		is included in Table \ref{tab:stars}, and is also plotted as a function of the wind momentum fluxes in 
		Figure~\ref{fig:diag}. Note that, for any pair of stars, $d$  changes with time as they orbit in the Galactic Centre, 
		therefore colliding wind stars can go through both regimes.
		Star 33E has the largest value of $d^{\rm cool}_*$. Encounters involving this star with similar ones at 10 mpc-scales should 
		result in the formation of cold slabs. For the rest of the sample we see that only for  separations below $0.1-1$~mpc (20--206 AU)
		their winds would be radiatively efficient. Such short separations are in the range of close binary systems (Section~\ref{sec:binaries}).
	\subsection{Mass-losing stars encounters in the Galactic Centre}
	\label{sec:encon}
	
		Now we turn our attention to the stellar orbits in the Galactic centre, to study how often close encounters between single mass-losing stars are produced .
		To do so, we ran a simple, Newtonian gravity, test-particle simulation to follow 
		the stars around Sgr~A*.  This approximation is correct, as the distances from the stars to Sgr~A* are too large for relativistic 
		effects to be important, and the time-scale for stellar scattering to be relevant is much longer than the period we are interested in 
		\citep[see][]{Alexander05}. We ran the models for $10^4\,$yr, which is already a much longer timescale than the expected lifetime 
		of a cloud like G2 \citep{B12}.
		
		We only registered encounters of stellar pairs at distances shorter than 10 mpc (2060 AU) based on our parameter space study, noting the 
		minimum distance of the close passage $d_{\rm min}$ and the duration of the encounter (i.e. the time stars are closer than 10 mpc). 
		As initial conditions we used the 3D velocities and 2D sky positions observed by \cite{P06}, meanwhile the unobservable $z$-coordinate 
		was chosen using different assumptions for the orbital distribution. We used the 3 different models from the work by \cite{C08} 
		to obtain the $z$-coordinate of our young star sample: \textit{Min-ecc}, \textit{1Disc} and \textit{2Discs}. \textit{Min-ecc} uses the 
		$z$-coordinate values that minimise the orbital eccentricities, \textit{1Disc} assumes roughly half of the stars are in the well defined 
		clockwise disc \citep{B06}, and \textit{2Discs} assumes the existence of both clockwise and counter-clockwise discs. The results of 
		this procedure are summarised in Table \ref{tab:encon},  where we see that over a $10^4\,$yr period, there is at most one encounter 
		with a $\sim1$ mpc separation\footnote{Notice that this ``encounter" happens between stars 13E2 and 13E4, which might be bound 
		together by a dark mass \citep{Fritz10}, not included in our calculations. That dark mass would, however, likely increase the encounter 
		duration.}. From the typical encounter duration of $\sim50$~yr and our integration time, we can estimate a rough probability of 
		producing G2 through encounters of single stars as 0.5\%.
		\begin{table*}
			\centering
  			\caption{Galactic Centre mass-losing star encounters at $< 10$~mpc (2060~AU). Column 1: initial conditions model name. Column 2: IDs of the stars in the encounter. 
			For each pair, the asterisk symbol (*) marks the star with the weaker wind (i.e., smaller momentum flux). When no asterisk is shown, both stars have the same 
			momentum flux. Columns 3 and 4: minimum stellar separation in the encounter in mpc and AU, respectively. Column 5: time stars are closer than 10~mpc (2060~AU). 
			Column 6: wind momentum flux ratio estimated from Equation~\ref{eq:eta}. Columns 7 and 8: CD distance to the weaker wind star calculated from Equation~\ref{eq:r2} in 
			mpc and AU, respectively. Column 9: weaker wind cooling parameter obtained from Equation~\ref{eq:chiasym}. Column 10: whether the weaker wind is radiative or not 
			at the minimum stellar separation of the encounter.}
			\label{tab:encon}
  			\begin{tabular}{@{}ccrrrcrrcc@{}}
				\hline
				Model	&	Stars			&	\multicolumn{2}{c}{$d_{\rm min}$}	&	Duration	&	$\eta$	&	\multicolumn{2}{c}{$R_2$}	&	Weaker wind 		&	Thin shell?\\
						&				&	(mpc)	&	(AU)				&	(yr)		&			&	(mpc)	&	(AU)		&	$\chi_{\rm asym}$	&	($\chi_{\rm asym}<1$)\\
				\hline
				\hline
				\textit{1Disc}&				&			&					&			&			&			&			&					&\\
						&	40 -- 60$^*$	&	9.2		&	1895				&	10		&	0.021	&	1.16		&	239		&	7.581			&	NO	\\
						&	48 -- 60$^*$	&	4.0		&	824				&	80		&	0.034	&	0.62		&	128		&	4.052			&	NO	\\
						&	48 -- 51$^*$	&	1.0		&	206				&	1200		&	0.304	&	0.36		&	74		&	0.260			&	YES	\\
						&	51 -- 60$^*$	&	6.0		&	1236				&	70		&	0.112	&	1.50		&	309		&	9.797			&	NO	\\
				\hline
				\textit{2Discs}&				&			&					&			&			&			&			&					&\\
						&	19$^*$ -- 48	&	3.0		&	618				&	15		&	0.061	&	0.59		&	122		&	0.519			&	YES	\\
						&	19$^*$ -- 51	&	7.0		&	1442				&	20		&	0.200	&	2.16		&	445		&	1.889			&	NO	\\
						&	19$^*$ -- 48	&	2.0		&	412				&	220		&	0.061	&	0.40		&	82		&	0.347			&	YES	\\
						&	19$^*$ -- 51	&	4.0		&	824				&	20		&	0.200	&	1.24		&	255		&	1.080			&	NO	\\
						&	19$^*$ -- 51	&	9.5		&	1957				&	10		&	0.200	&	2.94		&	606		&	2.564			&	NO	\\
						&	20 -- 32$^*$	&	8.0 		&	1648				&	60		&	0.776	&	3.75		&	773		&	51.16			&	NO	\\
						&	31 -- 72$$		&	7.5 		&	1545				&	20		&	1.000	&	3.75		&	773		&	51.20			&	NO	\\
						&	39$^{*}$ -- 51	&	2.0 		&	412				&	15		&	0.434	&	0.79		&	163		&	0.534			&	YES	\\
						&	41$^{*}$ -- 48	&	9.0 		&	1854				&	10		&	0.065	&	1.83		&	377		&	0.240			&	YES	\\
						&	48 -- 51$^{*}$	&	5.0 		&	1030				&	550		&	0.304	&	1.78		&	367		&	1.297			&	NO	\\
				\hline
				\textit{Min-ecc}&			&			&					&			&			&			&			&					&		\\
						&	23$^*$ -- 41	&	8.0 		&	1648				&	50		&	0.945	&	3.94		&	812		&	3.442			&	NO	\\
						&	48 -- 51$^*$	&	1.0 		&	206				&	1200		&	0.304	&	0.36		&	74		&	0.260			&	YES	\\
						&	48 -- 60$^*$	&	4.0 		&	824				&	80		&	0.034	&	0.62		&	128		&	4.058			&	NO	\\
						&	48 -- 61$^*$	&	6.0 		&	1236				&	20		&	0.034	&	0.93		&	192		&	6.084			&	NO	\\
						&	51 -- 60$^*$	&	6.0 		&	1236				&	70		&	0.112	&	1.50		&	309		&	9.797			&	NO	\\
						&	51 -- 61$^*$	&	7.5 		&	1545				&	25		&	0.112	&	1.88		&	387		&	12.25			&	NO	\\
				\hline
			\end{tabular}
		\end{table*}
	\subsection{Asymmetric mass losing stars encounters}
	\label{sec:asymm}
		For simplicity, we have only analysed collisions of identical stellar winds. However, in reality we will typically have 
		encounters between stars with different wind properties. To study such cases, detailed numerical simulations are needed,
		such as those performed by \cite{P09,V10,L11}.
		These authors have shown that even a small velocity difference in the colliding winds can excite the KHI. 
		This would mix two-phase material on top of other instabilities that can take place simultaneously. Although all these processes can be 
		very complicated to track analytically, based on our work we can give a qualitative description of possible scenarios that 
		can take place for asymmetric close encounters. 
		Stellar wind encounters are characterised by their momentum flux ratio \citep{L90},
		\begin{equation}
			\eta = \frac{\dot{M}_2V_{2}}{\dot{M}_1V_1},
			\label{eq:eta}
		\end{equation}
		where the subscript $2$ stands for the weaker wind and $1$ for the stronger one, so $\eta\le1$ by definition. 
		Notice that $\eta$ is independent of the stellar separation, provided this is large enough to allow the winds to reach their terminal velocities.
		For $\eta\neq1$, the interaction zone of the shocked gas bends towards the weaker star. The CD distance to the 
		weaker wind star will be given by
		\begin{equation}
			R_2	= \frac{\sqrt{\eta}}{1+\sqrt{\eta}}d.
			\label{eq:r2}
		\end{equation}
		Then, the CD will be located closer to the weaker star for systems with smaller $\eta$. Furthermore, 
		the weaker wind will be less diluted before the collision producing a denser slab (compared to systems with 
		same $\eta$ but different stellar separation). In this way, stars with large momentum fluxes in their winds can 
		be important in asymmetric encounters as they could ``force" weaker winds to radiate their energy rapidly, which would result in radiative shocks.
		In Figure \ref{fig:diag}, we highlighted the stars with largest momentum fluxes in their winds, including their names on the plot.
		Encounters of these stars with others, of the left side of the plot, will produce encounters with small $\eta$ and short distances 
		$R_2$ from the weaker star to the CD.
		For each of the encounters we registered in the previous subsection, we calculated the momentum flux ratio and the $R_2$ value. 
		We can now modify the definition of the cooling parameter (Equation~\ref{eq:chi}), so it uses the distance $R_2$  at which the CD is 
		expected to form in the asymmetric close encounters,
		\begin{equation}
			\chi_{\rm asym} \approx \frac{1}{2}\frac{V_8^{5.4}R_{2,12}}{\dot{M}_{-7}},
			\label{eq:chiasym}
		\end{equation} 
		where $R_{2,12} = R_{2}/10^{12}\,$cm. Since $R_2$ is always a fraction of $d/2$, the density of the weaker wind will be higher 
		at that position compared to the symmetric case and the slab will be able to radiate its energy away more rapidly. All these estimates 
		are included in Table~\ref{tab:encon}. From those results we can check that for two of our models (\textit{1Disc} and \textit{Min-ecc}) 
		there is only one encounter where the weaker wind produces a cold slab that might become unstable, while in the other 
		case (\textit{2Discs}), there are four such encounters. The difference is at least partially due to the fact that in the 
		\textit{2Discs} case the stars are closer together and more encounters are produced in general. These systems deserve 
		more study because they could be clump sources.
		It is important to remark that, in the asymmetric case, on one side of the CD we have a thin shell while on the other we have the 
		hot shocked gas of the stronger wind. The latter tends to stabilise any instability possibly excited, so even if $\chi_{\rm asym}<1$ 
		there might be no clump formation. Also notice that all these estimates were done under the assumption that the stars are separated 
		well enough in order to accelerate their winds up to their terminal velocities.  However, for extreme values of $\eta$ that might 
		not be the case.  		
\section{Discussion}
	\label{sec:discussion}
	\subsection{Limitations and uncertainties in the model}
		Although our model has been very useful to test the likelihood of clump formation in stellar wind shocked gas, there are two important
		assumptions we have to consider before comparing with other works, specially with 2- or 3-dimensional models.
		\subsubsection{The planar winds assumption}
		\label{sec:planar}
			Our model considers planar shocks, rather than more realistic spherical wind collisions.
			A planar shock geometry is a good approximation for the wind collision at the apex (i.e., at the CD and two-star axis intersection), 
			where the gas effectively moves along the axis joining both stars. 
			Off this axis, the winds will have a perpendicular velocity component, and the planar approximation breaks down.  
			Moreover, the off-axis velocity will prevent the material from accumulating in the slab. 
			Nevertheless, the well-modelled apex is arguably the most interesting location of the wind collision for us, as this is where we expect the most massive clumps to form. 
			As shown in Figure~\ref{fig:mass-range}, as long as we stay in the radiatively-efficient regime, higher values of $\chi$ generate more massive clumps. 
			So far, we have defined $\chi$ using the wind speed and the distance between the stars, which is only appropriate along the two-stars axis. 
			In a more realistic model, $\chi$ changes along the CD, decreasing from the apex as we show below.
			
			Let us define a more general cooling parameter as $\chi$~=~$\chi(\theta)$, where $\theta$ is the angle between the two-star axis and a line connecting 
			one star and an arbitrary point $P$ on the CD. The distance from the star to $P$ will be given by $d_{*}'$~=~$d_{*}/\cos\theta$. 
			Furthermore, the shock at $P$ will be weaker than at the apex, as only the component perpendicular to the CD will contribute to it. 
			This component will be given by $V'=V\cos\theta$. Using these quantities we can generalise Equation~\ref{eq:chi} obtaining
			\begin{equation}
				\chi(\theta)=\frac{1}{2}\frac{V_8^{5.4}d_{*12}}{\dot{M}_{-7}}(\cos\theta)^{4.4} = \chi_{\perp}(\cos\theta)^{4.4},
			\end{equation}
			\noindent where $\chi_{\perp}$ is the cooling parameter at the apex, as previously defined in Equation~\ref{eq:chi}. 
			From this new expression we see that in general, $\chi(\theta)\leq\chi_{\perp}$, recovering Equation~\ref{eq:chi} when $\theta=0$ as expected. 
			Thus, away from the apex the cooling parameter decreases, and according to our results, we expect the creation of less massive clumps. 
			With this in mind, the clump masses we quote are the largest we can expect from a given wind collision and they would form near the apex where the planar-wind assumption is accurate.
		\subsubsection{Impact of metallicity on radiative cooling} 
		\label{sec:limit}
			Different metallicities result in different cooling timescales, which in turn modify the sizes 
			and masses of the clumps. As there is no agreement on the metallicity measurements for the massive stars in the Galactic Centre \citep[see the review by][]{Genzel10} 
			we have set $Z=3Z_{\odot}$, following \cite{C05} and close to the values \cite{M07} studied. However, in order to quantify the sensitivity to our choice, 
			we also produced models with metallicities of $Z_{\odot}$ and $5Z_{\odot}$ in the analytical cooling function we use (see Equation~\ref{eq:dis}). 
			We found changes of a factor  $\sim2$ in the clump sizes, which translate in a factor $\sim8$ for the clump masses. 
			As expected, a lower metallicity value increases the cooling timescale, and viceversa. Nevertheless, these changes in the metallicity do not result in 
			switching the wind regimes for the system we have analysed (e.g., from radiative to adiabatic or in the opposite direction) as the impact on the cooling 
			parameter is not very strong for the metallicity values we have tested. 
			Thus, we are confident our results do not depend strongly on our chosen metallicity. Still, one should be cautious when 
			comparing with other works which could have used other metallicity values, and therefore another cooling function.
	\subsection{Binary stars}
	\label{sec:binaries}
		
		Based on our results, clump formation seems not very likely to occur in the environment of the Galactic Centre, due to 
		either symmetric or asymmetric stellar encounters. However, we have not studied a case that probably deserves more attention: 
		\textit{colliding winds binary systems}.
		
		As shown in Section~\ref{sec:diagn}, the formation of cold slabs typically requires stellar separations below 1 mpc.  While those
		separations are not often achieved in stellar encounters of single stars, they are easily reached by close binaries. The census of 
		young massive binary stars in the Galactic Centre is still  incomplete, but the recent study by \cite{P14} increased to three the 
		amount of confirmed binary systems:$(i)$ IRS 16SW, a 19.5-day period Ofpe/WN9 eclipsing contact binary \citep{M06}, $(ii)$ 
		IRS 16NE, a 224-day period Ofpe/WN9 binary and $(iii)$ E60\footnote{The star with ID 60 in Table~\ref{tab:stars}.}, a 2.3-day 
		eclipsing contact Wolf-Rayet binary. The inferred separations of these binary systems are of the order of 10 $\mu$pc and below. 
		That, together with the wind properties listed in Table~\ref{tab:stars}, could make these three binaries very effective sites for clump 
		formation.  IRS 16SW is of particular interest, as it has a clockwise orbit that roughly coincided with G2's at the latter's apocentre. 
		Notice that the wind properties of these stars were assigned by \cite{C08} as mentioned in Section~\ref{sec:colliding}.
		However, we would not expect any important wind component to be substantially faster, as it would show up as broader lines in the 
		spectra. Additionally, from the estimations by \cite{P14} we expect an overall $\sim 30$\% spectroscopic binary fraction for the 
		massive OB/WR stars in the Galactic Centre. Thus, even though  our current knowledge of the binary population in the Galactic 
		Centre is limited, close binary systems remain as a very promising possibility for the creation of cold clumps. A more detailed 
		numerical model of this process is deferred to a forthcoming paper.
\section{Conclusions}
	\label{sec:conclusions}

	We have developed a simple and straightforward prescription to study clump formation through the NTSI 
	mechanism in symmetric colliding wind systems. The input parameters are the mass loss rate, wind terminal 
	velocity and stellar separation. Radiatively efficient wind collisions are capable of creating cold gas clumps 
	($10^4$~K) in a wide range of masses, where the most massive ones are of the order of the Earth mass for 
	strong outflows ($\dot{M}\sim10^{-5}$~$\msun$~yr$^{-1}$), relatively slow wind terminal velocities ($250-750$~km~s$^{-1}$) 
	and short stellar separations ($0.1-10$~mpc or $20-2060$ AU).  Nevertheless, the wide range of unstable wavelengths that are excited 
	prevents us from predicting unequivocal clump masses, as shorter-wavelength perturbations grow faster and might 
	hinder the development of larger-scale ones.
		
	We also found that the possible  clump masses depend strongly on the timescale needed for the slab to collapse within 
	the radiative wind regime. The most massive ones would be generated in systems where the shocked gas does not cool 
	instantaneously, i.e. systems with $\chi$ approaching unity. Studying that regime however is not straightforward as both 
	radiative and adiabatic cooling are important. 
		
	Our results show that the formation of gas clumps with masses comparable to the G2 cloud is indeed possible in symmetric 
	colliding winds. However, this scenario does not seem likely in the Galactic Centre given the currently known mass-losing 
	star sample, as the required sub-mpc separations are very rarely achieved by them. We also discussed clump formation in 
	asymmetric encounters, finding that the massive and slow outflow of IRS 33E could create clumps if confined by a powerful 
	wind of another star. Similarly, the collision of winds from IRS 13E2 and 13E4 could also generate a cold slab unstable to the NTSI.  
	These stars have similar orbits and  spend a significant time at short separations, which could explain the presence of many 
	dusty clumps in their vicinity \citep[see][]{Fritz10}. However, they orbit Sgr~A* in the opposite sense as G2, making this pair 
	an unlikely origin for this particular cloud. A promising possibility is that clumps are produced in close binaries, of which three 
	are currently know. Still, better observational data are required to constrain the wind properties of both components of each binary.
	The IRS 16SW binary is of particular interest, as its orbit coincides with G2's at apocentre.
		
	In conclusion, given our current analysis and the available stellar wind data, the formation of G2-like clouds 
	in the Galactic centre appears as a possible but not very common event. We defer more concrete results to a 
	future study using 2D and 3D numerical modelling. This is required to treat systems with unequal stellar winds, 
	with $\chi\sim1$, or for winds that collide before reaching their terminal velocity, such as compact binaries. 
	Numerical models are also required to follow the growth of different unstable wavelengths and obtain a clump mass function. 
\section*{Acknowledgments}
	We thank the anonymous referee for very useful comments.  
	We also thank Cristian Hern\'andez for double checking some of the calculations. Part of this work was carried out at MPE, 
	which DC and JC thank for the warm hospitality. We acknowledge support from CONICYT--Chile through FONDECYT 
	(1141175), Basal (PFB0609) and Anillo (ACT1101) grants. DC is supported by CONICYT--PCHA/Doctorado Nacional (2015--21151574). 
	A. Ballone was supported by the Deutsche Forschungsgemeinschaft (DFG) priority program 1573 (Physics of the Interstellar Medium) 
	and by the DFG Cluster of Excellence ``Origin and Structure of the Universe".


\label{lastpage}

\end{document}